\documentstyle[prl,aps,preprint]{revtex}
\def\beginwide{
        \end{multicols} \vspace*{-0.5cm} \noindent
        \rule{3.5in}{.1mm}\rule{.1mm}{5mm} \widetext \medskip }
\def\endwide{
        \hspace*{3.5in}~\rule[-5mm]{.1mm}{5mm}\rule{3.5in}{.1mm}
        \begin{multicols}{2}\narrowtext \vspace*{-1.0cm} \noindent }
\def\beginwidetop{
        \end{multicols} \vspace*{-0.5cm} \noindent
        \widetext \medskip }
\def\endwidebottom{
        \begin{multicols}{2} \vspace*{-1.0cm} \noindent }
\def\sqr#1#2{{\vcenter{\hrule height.#2pt\hbox{\vrule width.#2pt height#1pt \kern#1pt \vrule width.#2pt}\hrule height.#2pt}}}
\def\square{\mathchoice\sqr64\sqr64\sqr{4.2}3\sqr{3.0}3}

\begin{document}

%\draft
\widetext

\title{Metric--affine gauge theory of gravity \\II. Exact solutions}
\author{ Friedrich W. Hehl\thanks{E-mail: hehl@thp.uni-koeln.de.
    Permanent address: Institute for Theoretical Physics, University
    of Cologne, D--50923 K\"oln, Germany} and Alfredo
  Mac\'{\i}as\thanks{E-mail: amac@xanum.uam.mx}\\ Departamento de
  F\'{\i}sica\\ Universidad Aut\'onoma Metropolitana--Iztapalapa\\ 
  Apartado Postal 55--534, C.P. 09340, M\'exico, D.F., Mexico\\ }

%\date{\today}

\maketitle

\begin{abstract}
  In continuing our series on metric-affine gravity (see Gronwald, IJMP
  {\bf D6} (1997) 263 for Part I), we review the exact solutions of
  this theory.  {\em file magexac7.tex, 1999-04-09}
\end{abstract}


%\begin{multicols}{2}

%**********************************************************

\section{Metric--affine gravity (MAG)}
                                     
In 1976, a new metric--affine theory of gravitation was published
\cite{MAG}.  In this model, the {\em metric} $g_{ij}$ and the linear
(sometimes also called affine) {\em connection} $\Gamma_{ij}{}^k$ were
considered to be independent gravitational field variables. The metric
carries 10 and the connection 64 independent components. Although
nowadays more general Lagrangians are considered, like the one in
Eq.(\ref{6}), the original Lagrangian density of metric--affine
gravity reads
\begin{equation}
  {\cal V}_{{\rm GR}^\prime} = \frac{\sqrt{-g}}{2\kappa}\, g^{ij}
  \Bigl[{\rm Ric}_{\,ij} (\Gamma, \partial \Gamma) + \beta\, Q_i Q_j
  \Bigr]
\label{1}\, .
\end{equation}
The Ricci tensor ${\rm Ric}_{\,ij}$ depends only on the connection but
not on the metric, whereas the Weyl covector $Q_i := -g^{kl}\,
\nabla_i\, g_{kl}/4$ depends on both. Here $\nabla_i$ represents the
covariant derivative with respect to the connection $\Gamma_{ij}{}^k$,
furthermore $g=\det\,g_{kl}$, $\kappa$ is Einstein's gravitational
constant, and $\beta$ a dimensionless coupling constant.  With $i,j,k,
\cdots=0,1,2,3$ we denote coordinates indices.

This model leads back to general relativity, as soon as the material
current coupled to the connection, namely $\sqrt{-g}\,
\Delta^{ij}{}_k:= \delta {\cal L}_{mat}/\delta \Gamma_{ij}{}^k$, the
so--called hypermomentum, vanishes. Thus, in such a model, the
post--Riemannian pieces of the connection and the corresponding new
interactions are tied to matter, they do not propagate.

As we know from the weak interaction, a contact interaction appears to
be suspicious for causality reasons, and one wants to make it {\em
  propagating}, even if the carrier of the interaction, the
intermediate gauge boson, may become very heavy as compared to the
mass of the proton, e.g..  However, before we report on the more
general gauge Lagrangians that have been used, we turn back to the
geometry of spacetime.

MAG represents a gauge theory of the 4--dimensional affine group
enriched by the existence of a metric. As a gauge theory, it finds its
appropriate form if expressed with respect to arbitrary frames or {\em
  coframes}.  Therefore, the apparatus of MAG was reformulated in the
calculus of exterior differential forms, the result of which can be
found in the review paper \cite{PartI}, see also \cite{PRs} and
\cite{Erice95}. Of course, MAG could have been alternatively
reformulated in tensor calculus by employing an arbitrary
(anholonomic) frame (tetrad or vierbein formalism), but exterior
calculus, basically in a version which was advanced by Trautman
\cite{Trautman} and others \cite{Jim,Chen+Jim}, seems to be more
compact.

In the new formalism, we have then the metric $g_{\alpha\beta}$, the
coframe $\vartheta^\alpha$, and the connection 1--form
$\Gamma_\alpha{}^\beta$ (with values in the Lie algebra of the
4--dimensional linear group $GL(4,R)$) as new independent field
variables. Here $\alpha,\beta,\gamma,\cdots = 0,1,2,3$ denote
(anholonomic) frame indices. For the formalism, including the
conventions, which we will be using in this paper, we refer to
\cite{PRs}.  

A first order Lagrangian formalism for a matter field $\Psi$ minimally
coupled to the gravitational {\em potentials} $g_{\alpha\beta}$,
$\vartheta^\alpha$, $\Gamma_\alpha{}^\beta$ has been set up in
\cite{PRs}.  Spacetime is described by a metric--affine geometry with
the gravitational {\em field strengths} nonmetricity $Q_{\alpha\beta}
:=-Dg_{\alpha\beta}$, torsion $T^\alpha:=D\vartheta^\alpha$, and
curvature $R_\alpha{}^\beta:= d\Gamma_\alpha{}^\beta
-\Gamma_\alpha{}^\gamma\wedge\Gamma_\gamma{}^\beta$.  The
gravitational field equations
\begin{eqnarray} 
  DH_{\alpha}- E_{\alpha}&=&\Sigma_{\alpha}\,,\label{first}\\ 
  DH^{\alpha}{}_{\beta}-
  E^{\alpha}{}_{\beta}&=&\Delta^{\alpha}{}_{\beta}\,,
\label{second}
\end{eqnarray} 
link the {\em material sources}, the material energy--momentum current
$\Sigma_\alpha$ and the material hypermomentum current
$\Delta^\alpha{}_\beta$, to the gauge field {\em excitations}
$H_\alpha$ and $H^\alpha{}_\beta$ in a Yang--Mills like manner. In
\cite{PRs} it is shown that the field equation corresponding to the
variable $g_{\alpha\beta}$ is redundant if (\ref{first}) as well as
(\ref{second}) are fulfilled.

If
the gauge Lagrangian 4--form
\begin{equation}
  V= V\left(g_{\alpha\beta}, \vartheta^\alpha, Q_{\alpha\beta},
    T^\alpha, R_\alpha{}^\beta \right)
\label{2}\,
\end{equation}                                        
is given, then the excitations can be calculated by partial
differentiation,
\begin{equation}  H_\alpha = - \frac{\partial V}{\partial T^\alpha}\, , \quad 
  H^\alpha{}_\beta= -\frac{ \partial V}{\partial
    R_\alpha{}^\beta}\,,\quad M^{\alpha\beta} = - 2 \frac{\partial
    V}{\partial Q_{\alpha\beta}}\, ,
\label{3}\, 
\end{equation}
whereas the gauge field currents of energy--momentum and
hypermomentum, respectively, turn out to be linear in the Lagrangian
and in the excitations,
\begin{eqnarray}
  E_{\alpha} & := & \frac{\partial V}{\partial\vartheta^\alpha}
  =e_{\alpha}\rfloor V + (e_{\alpha}\rfloor T^{\beta}) \wedge
  H_{\beta} + (e_{\alpha}\rfloor R_{\beta}{}^{\gamma})\wedge
  H^{\beta}{}_{\gamma} + {1\over 2}(e_{\alpha}\rfloor Q_{\beta\gamma})
  M^{\beta\gamma}\,,\\ E^{\alpha}{}_{\beta} & := &\frac{\partial
    V}{\partial\Gamma_\alpha{}^\beta}= - \vartheta^{\alpha}\wedge
  H_{\beta} - g_{\beta\gamma}M^{\alpha\gamma}\,.
\end{eqnarray}
Here $e_\alpha$ represents the frame and $\rfloor$ the interior
product sign, for details see \cite{PRs}.

%***********************************************
\section{The quadratic gauge Lagrangian of MAG}

The gauge Lagrangian (\ref{1}), in the new formalism, is a 4--form and 
reads \cite{Mapping}
\begin{equation}
  V_{{\rm GR}^\prime} = \frac{1}{2\kappa} \left( - R^{\alpha\beta}
    \wedge \eta_{\alpha\beta} + \beta Q \wedge {}^*Q\right)
\label{4}\, .
\end{equation}
Here $\eta_{\alpha\beta} := {}^*(\vartheta_\alpha \wedge
\vartheta_\beta)$, $*$ denotes the Hodge star. Besides Einstein
gravity, it encompasses additionally {\em contact} interactions.

It is obvious of how to make $Q$ a propagating field: One adds, to the
massive $\beta$--term, a kinetic term \cite{HLordSmalley,PonoObukhov}
$-\alpha\,dQ\wedge {}^*dQ/2$. Since $dQ= R_\gamma{}^\gamma/2$, the
kinetic term can alternatively be written as
\begin{equation}
-\frac{\alpha}{8} \,R_\beta{}^\beta \wedge {}^*R_\gamma{}^\gamma
\label{5}\, .
\end{equation}
This term, with the appearance of one Hodge star, displays a typical
{\em Yang--Mills structure}. More generally, propagating
post--Riemannian gauge interactions in MAG can be consistently
constructed by adding terms quadratic in $Q_{\alpha\beta}$,
$T^\alpha$, $R_\alpha{}^\beta$ to the Hilbert-Einstein type Lagrangian
and the term with the cosmological constant.

In the first order formalism we are using, higher order terms, i.e.\
cubic and quartic ones etc.\ would preserve the second order of the
field equations. However, the {\em quasilinearity} of the gauge field
equations would be destroyed and, in turn, the Cauchy problem would be
expected to be ill--posed. Therefore we do not go beyond a gauge
Lagrangian which is quadratic in the gauge field strengths
$Q_{\alpha\beta}$, $T^\alpha$, $R_\alpha{}^\beta$. Incidentally, a
quadratic Lagrangian is already so messy that it would be hard to
handle a still more complex one anyway.

Different groups have already added, within a metric--affine
framework, different quadratic pieces to the Hilbert--Einstein--type
Lagrangian, see
\cite{Yasskin,Grossmann4,Duan,irredDermott,TresguerresShear1,TuckerWang,TuckWar,Teyssandier,Yuritheorem,collwavesMAG},
e.g., and references given there.  The end result of all these
deliberations is the {\em most general parity conserving quadratic}
Lagrangian which is expressed in terms of the $4+3+11$ irreducible
pieces (see \cite{PRs}) of $Q_{\alpha\beta}$, $T^\alpha$,
$R_\alpha{}^\beta$, respectively:
\begin{eqnarray} 
\label{QMA} V_{\rm MAG}&=&
\frac{1}{2\kappa}\,\left[-a_0\,R^{\alpha\beta}\wedge\eta_{\alpha\beta}
  -2\lambda\,\eta+T^\alpha\wedge{}^*\!\left(\sum_{I=1}^{3}a_{I}\,^{(I)}
    T_\alpha\right)\right.\nonumber\\ &+&\left.
  2\left(\sum_{I=2}^{4}c_{I}\,^{(I)}Q_{\alpha\beta}\right)
  \wedge\vartheta^\alpha\wedge{}^*\!\, T^\beta + Q_{\alpha\beta}
  \wedge{}^*\!\left(\sum_{I=1}^{4}b_{I}\,^{(I)}Q^{\alpha\beta}\right)\right.
\nonumber \\&+&
b_5\bigg.\left(^{(3)}Q_{\alpha\gamma}\wedge\vartheta^\alpha\right)\wedge
{}^*\!\left(^{(4)}Q^{\beta\gamma}\wedge\vartheta_\beta \right)\bigg]
\nonumber\\&- &\frac{1}{2\rho}\,R^{\alpha\beta} \wedge{}^*\!
\left(\sum_{I=1}^{6}w_{I}\,^{(I)}W_{\alpha\beta}
  +w_7\,\vartheta_\alpha\wedge(e_\gamma\rfloor
  ^{(5)}W^\gamma{}_{\beta} ) \nonumber\right.\\&+& \left.
  \sum_{I=1}^{5}{z}_{I}\,^{(I)}Z_{\alpha\beta}+z_6\,\vartheta_\gamma\wedge
  (e_\alpha\rfloor ^{(2)}Z^\gamma{}_{\beta}
  )+\sum_{I=7}^{9}z_I\,\vartheta_\alpha\wedge(e_\gamma\rfloor
  ^{(I-4)}Z^\gamma{}_{\beta} )\right)
\label{6}\,.  
\end{eqnarray}
The constant $\lambda$ is the cosmological constant, $\rho$ the strong
gravity coupling constant, the constants $ a_0, \ldots a_3$, $b_1,
\ldots b_5$, $c_2, c_3,c_4$, $w_1, \ldots w_7$, $z_1, \ldots z_9$ are
dimensionless. We have introduced in the curvature square term the
irreducible pieces of the antisymmetric part $W_{\alpha\beta}:=
R_{[\alpha\beta]}$ and the symmetric part $Z_{\alpha\beta}:=
R_{(\alpha\beta)}$ of the curvature 2--form.  In $Z_{\alpha\beta}$, we
have the purely {\em post}--Riemannian part of the curvature. Note the
peculiar cross terms with $c_I$ and $b_5$.

Esser \cite{Esser}, in the component formalism, has carefully
enumerated all different pieces of a quadratic MAG Lagrangian, for the
corresponding nonmetricity and torsion pieces, see also Duan et al.\ 
\cite{Duan}.  Accordingly, Eq.(\ref{6}) represents the most general
quadratic parity--conserving MAG--Lagrangian. All previously published
quadratic parity--conserving Lagrangians are subcases of (\ref{6}).
Hence (\ref{6}) is a safe starting point for our future
considerations.

We concentrate here on Yang--Mills type Lagrangians. Since $V_{\rm
  MAG}$ is required to be an {\em odd} 4--form, if parity conservation
is assumed, we have to build it up according to the scheme $F\wedge
{}^*F$, i.e.\ with one Hodge star, since the star itself is an odd
operator. Also the Hilbert--Einstein type term is of this type, namely
$\sim R^{\alpha\beta} \wedge {}^*(\vartheta_\alpha \wedge
\vartheta_\beta)$, as well as the cosmological term $\sim \eta =
{}^*1$. Thus $V_{\rm MAG}$ is homogeneous of order one in the star
operator. It is conceivable that in future one may want also consider
parity violating terms with no star appearing (or an even number of
them) of the (Pontrjagin) type $F \wedge F$. Typical terms of this
kind in four dimensions would be
\begin{equation}
  R^{\alpha\beta} \wedge (\vartheta_\alpha \wedge \vartheta_\beta)\,
  ,\quad 1\, , \quad T^\alpha\wedge T_\alpha\, ,\quad Q_{\alpha\beta}
  \wedge\vartheta^\alpha \wedge T^\beta\,,\quad R^{\alpha\beta} \wedge
  R_{\alpha\beta}
\label{7} \, .
\end{equation}   
The first term of (\ref{7}), e.g., represents the totally
antisymmetric piece of the curvature $R^{[\gamma\delta\alpha\beta]}\,
\vartheta_\gamma\wedge\vartheta_\delta \wedge\vartheta_\alpha\wedge
\vartheta_\beta$, which is purely post--Riemannian.  Such
parity--violating Lagrangians have been studied in the past, see,
e.g., \cite{Mukku,Nelson} and \cite{Bianchi,O+H}, but, for simplicity,
we will restrict ourselves in this article to parity preserving
Lagrangians.

%*************************************************
\section{On the possible physics of MAG} 

Here we are, with a Lagrangian $V_{\rm MAG}$ encompassing more than
two dozens of unknown dimensionless constants.  But the situation is
not as bad as it may look at first. For the Newton--Einstein type of
{\em weak gravity} --- the corresponding terms are collected in
(\ref{6}) within two square brackets $[ \quad ]$ --- we have the
gravitational constant $\kappa$, with dimension of $\kappa=length^2$,
and the cosmological constant $\lambda$, with dimension of $\lambda=
length^{-2}$. For {\em strong gravity} of the Yang--Mills type, the
basic newly postulated interaction within the MAG framework, the
strength of the coupling is determined by the dimensionless strong
coupling constant $\rho$. Thus, the three constants
$\kappa,\lambda,\rho$ are fundamental, whereas the rest of the
constants, 12 for weak and 16 for strong gravity, are expected to be
of the order unity or should partially vanish.

As was argued elsewhere \cite{PRs}, we do not believe that at the
present state of the universe the geometry of spacetime is described
by a metric--affine one. We rather think, and there is good
experimental evidence, that the present-day geometry is
metric-compatible, i.e., its nonmetricity vanishes. In earlier epochs
of the universe, however, when the energies of the cosmic ``fluid''
were much higher than today, we expect scale invariance to prevail ---
and the canonical dilation (or scale) current of matter, the trace of
the hypermomentum current $\Delta^\gamma{}_\gamma$, is coupled,
according to MAG, to the Weyl covector $Q^\gamma{}_\gamma$. By the
same token, shear type excitations of the material multispinors (Regge
trajectory type of constructs) are expected to arise, thereby
liberating the (metric-compatible) Riemann-Cartan spacetime from its
constraint of vanishing nonmetricity $Q_{\alpha\beta}=0$ . Tresguerres
\cite{Tres3} has proposed a simple cosmological model of Friedmann
type which carries a metric-affine geometry at the beginning of the
universe, the nonmetricity of which dies out exponentially in time.
That is the kind of thing we expect.

If one keeps the differential manifold structure of spacetime intact,
i.e., doesn't turn to discrete structures or non-commutative geometry,
then MAG appears to be the most natural extension of Einstein's
gravitational theory. The {\em rigid} metric-affine structure
underlying the Minkowski space of special relativity, see Kopczy\'nski
and Trautman \cite{K+T}, make us believe that this structure should
be gauged according to recipes known from gauge theory. Also the
existence, besides the energy-momentum current, of the {\em external}
material currents of spin and dilation (and, perhaps, of shear) does
point in the same direction.

\section{Exact MAG solutions of Tresguerres and Tucker \& Wang}

For getting a deeper understanding of the meaning and the possible
consequences of MAG, a search for exact solutions appears
indispensable. Tresguerres, after finding exact solutions
\cite{Tresguerres3D,Tres3Da} for specific $(1+2)$--dimensional models
of MAG, turned his attention to $1+3$ dimensions and, in 1994, for a
fairly general subclass of the Lagrangian (\ref{6}), found the first
static spherically symmetric solutions with a non--vanishing {\em
  shear charge} \cite{TresguerresShear1,TresguerresShear2}, i.e., the
solution is endowed with a traceless part
${\nearrow\!\!\!\!\!\!\!Q}_{\alpha\beta}:=
Q_{\alpha\beta}-Qg_{\alpha\beta}$ of the nonmetricity.  This
constituted a breakthrough. Since that time,
${\nearrow\!\!\!\!\!\!\!Q}_{\alpha\beta}$ lost its somewhat elusive
and abstract character. Even an operational interpretation has been
attempted in the meantime \cite{Test}.

The metric of Tresguerres' solution is the {\em Reissner--Nordstr\"om
  metric} of general relativity with cosmological constant but the
place of the electric charge is taken by the {\em dilation charge}
which is related to the trace of the nonmetricity, the Weyl covector.
Furthermore, the Tresguerres solutions carries, besides the
above-mentioned shear charge (related to the 

\begin{table}[h]
\caption{Irreducible decomposition of the nonmetricity$^*$ {\tt nom}
$Q_{\alpha\beta}$}
  \begin{center}
    \leavevmode
    
\begin{tabular}{|l|c|l|}
\hline
name& number of indep. comp.& \hfil piece   \hfil\\
\hline
\hline
{\tt nom}          & 40 & \hfil $Q_{\alpha\beta}$ \hfil\\
\hline
{\tt trinom}       & 16 & $^{(1)}Q_{\alpha\beta}:=Q_{\alpha\beta}
                                        -{}^{(2)}Q_{\alpha\beta}
                                        -{}^{(3)}Q_{\alpha\beta}
                                        -{}^{(4)}Q_{\alpha\beta}$\\
{\tt binom}        & 16 & ${}^{(2)}Q_{\alpha\beta}:={2\over3}
                           \,{}^*\!(\vartheta_{(\alpha}\wedge
                           \Omega_{\beta)})$\\
{\tt vecnom}       &  4 & ${}^{(3)}Q_{\alpha\beta}:={4\over 9}
                           \left(\vartheta_{(\alpha}e_{\beta)}
                             \rfloor\Lambda - {1\over
                               4}g_{\alpha\beta}\Lambda\right)$\\
{\tt conom}        &  4 & ${}^{(4)}Q_{\alpha\beta}:=g_{\alpha\beta}Q$\\
\hline
\end{tabular}

  \end{center}
\end{table}
\footnotesize
\noindent $^*)$ First 
the nonmetricity is split into its trace, the Weyl covector
$Q:={1\over 4}g^{\alpha\beta}Q_{\alpha\beta}$, and its traceless piece
${\nearrow\!\!\!\!\!\!\!Q}_{\alpha\beta}:=Q_{\alpha\beta}-
Qg_{\alpha\beta}$. The traceless piece yields the shear covector
$\Lambda:=\vartheta^{\alpha}e^{\beta}\rfloor
{\nearrow\!\!\!\!\!\!\!Q}_{\alpha\beta}$ and the shear 2-form
$\Omega_{\alpha}:=\Theta_{\alpha} - {1\over 3}e_{\alpha}\rfloor
(\vartheta^{\beta}\wedge\Theta_{\beta})$, with $\Theta_{\alpha}:=
{}^*({\nearrow\!\!\!\!\!\!\!Q}_{\alpha\beta}\wedge\vartheta^{\beta})$.
The 2-form $\Omega^{\alpha}$ describes ${}^{(2)}Q_{\alpha\beta}$ and
has precisely the same symmetry properties as the 2-form
${}^{(1)}T^{\alpha}$ (see below).  In particular, we can prove that
$e_{\alpha}\rfloor\Omega^{\alpha}=0$ and
$\vartheta_{\alpha}\wedge\Omega^{\alpha}=0$.  
\normalsize

%***************************************************************************

\begin{table}[h]
  \caption{Irreducible decomposition of the torsion$^{**}$ 
{\tt tor} $T^\alpha$}
  \begin{center}
    \leavevmode
\begin{tabular}{|l|c|l|}
\hline
name         & number of indep. comp.  & \hfil piece \hfil\\
\hline
\hline
{\tt tor}    & 24 & \hfil  $T^\alpha$ \hfil \\
\hline
{\tt tentor} & 16 & ${}^{(1)}T^{\alpha}:=T^{\alpha}
                     -{}^{(2)}T^{\alpha} - {}^{(3)}T^{\alpha}$\\
{\tt trator} &  4 & ${}^{(2)}T^{\alpha}:=  {1\over 3}
                     \vartheta^{\alpha}\wedge T$\\
{\tt axitor} &  4 & ${}^{(3)}T^{\alpha}:=-\,{1\over 3}
                     {}^*(\vartheta^{\alpha}\wedge A)$\\
\hline
\end{tabular}

  \end{center}
\end{table}
\footnotesize
\noindent $^{**})$ The 1-forms $T$ (torsion trace or covector) and $A$ (axial
covector) are defined by $T:=e_{\alpha}\rfloor T^{\alpha}$ and
$A:={}^*(\vartheta_{\alpha}\wedge T^{\alpha})$, respectively.
\normalsize

\clearpage

\noindent  traceless part of the
nonmetricity) a {\em spin charge} related to the torsion of spacetime.
Thus, beyond the Reissner--Nordstr\"om metric, the following
post--Riemannian degrees of freedom are excited in the Tresguerres
solutions (see Tables I and II): {\em two} pieces of the nonmetricity,
namely ${}^{(4)}Q_{\alpha\beta}$ ({\tt conom}, which is equivalent to
the Weyl covector) and the traceless piece ${}^{(2)}Q_{\alpha\beta}$
({\tt binom}), and all {\em three} pieces of the torsion
$^{(1)}T^\alpha$ ({\tt tentor}), $^{(2)}T^\alpha$ ({\tt trator}),
$^{(3)}T^\alpha$ ({\tt axitor}).  The names in the parentheses are
taken form our computer programs \cite{PRs,CPC}.  The first solution
\cite{TresguerresShear1}, requires in the Lagrangian weak gravity
terms and, for strong gravity, the curvature square pieces with $z_4
\neq 0$, $w_3 \neq 0$, $w_5 \neq 0$, i.e., with Weyl's segmental
curvature ({\tt dilcurv}), the curvature pseudoscalar ({\tt pscalar}),
and the antisymmetric Ricci ({\tt ricanti}). In his second solution
\cite{TresguerresShear2}, the torsion is independent of the
nonmetricity, otherwise the situation is similar yet not as clear cut.

The price Tresguerres had to pay in order to find exact solutions at
all was to impose {\em constraints} on the dimensionless {\em coupling
  constants} of MAG. In other words, the Lagrangian $V_{\rm MAG}$ was
engineered such that exact solutions emerged. This is, of course, not
exactly what one really wants.  Rather one would like to prescribe a
Lagrangian and then to find an exact solution. But, with the methods
then available, one could not do better.  And one was happy to find
exact solutions at all for such complicated Lagrangians.

As to the methods applied, one fact should be stressed. To handle
Lagrangians like (\ref{6}), it is practically indispensable to use
{\em computer algebra} tools. This is also what Tresguerres did. He
took Schr\"ufer's {\em Excalc} package of Hearn's computer algebra
system {\em Reduce;} for introductory lectures on Reduce and Excalc
see \cite{Stauffer}.  More recently, we described the corresponding
computer routines within MAG in some detail \cite{CPC} and showed of
how to build up Excalc programs for finding exact solutions of MAG.
What one basically does with these programs, is to make a clever
ansatz for the coframe, the torsion, and the nonmetricity, then to
substitute this into the field equations, as programed in Excalc, and
subsequently to inspect these expressions in order to get an idea of
how to solve them. One way of reducing them to a manageable size, is
to constrain the dimensionless coupling constants or to solve, also by
computer algebra methods, some of the partial differential equations
emerging. If \clearpage

\begin{table}[h]
\caption{Solutions for insular objects$^{*}$}
\bigskip
  
\begin{tabular}{|p{8cm}|c|p{5cm}|}
\hline
solution     &  references  &    post-Riemannian structures\\
\hline
\hline
{\bf Monopoles} with strong gravito--electric and strong gravito--magnetic 
charge (and combinations of them) plus triplet (degenerate case 
of Reissner-Nordstr\"om solution with triplet)
& \cite{Soliton2,M+S} & {\tt conom} $\sim$ {\tt vecnom} $\sim$
{\tt trator}\\\hline\hline
{\bf Reissner--Nordstr\"om} metric with strong gravito-electric charge plus
{\tt nom} and {\tt tor} 
&&\\\hline
--- dilation type solution  &\cite{TresguerresShear1,TuckerWang,Ho}&
{\tt conom} $\sim$ {\tt trator}, {\tt axitor} \cite{Ho}\\\hline
--- triplet type solution  &\cite{OVETH,Dereli}& {\tt conom} $\sim$
{\tt vecnom} $\sim$ {\tt trator} \\\hline
--- dilation--shear type solution &\cite{TresguerresShear1} &{\tt
  conom} $\sim$ {\tt binom},  {\tt tentor} $\sim$ {\tt binom}, {\tt
  trator} $\sim$ {\tt conom}, {\tt axitor} $\sim$ {\tt binom}\\\hline
--- dilation--shear--torsion type solution&\cite{TresguerresShear2}&
{\tt conom} $\sim$ {\tt binom}, {\tt tentor}, {\tt trator} $\sim$
{\tt conom}, {\tt axitor}\\\hline\hline
{\bf Kerr--Newman} metric with strong-gravito electric charge plus 
{\tt nom} and {\tt tor}&&\\\hline
--- triplet type solution& \cite{VTOH} &{\tt conom} $\sim$ {\tt
vecnom} $\sim$ {\tt trator}\\\hline\hline
{\bf Pleba\'nski--Demia\'nski} metric with strong gravito-electric and 
magnetic charge plus {\tt nom} and {\tt tor}&&\\\hline
--- triplet type solution &\cite{PDMAG}& {\tt conom} $\sim$ {\tt
vecnom} $\sim$ {\tt trator}\\\hline\hline 
{\bf Electrically} (and magnetically) charged versions of
all of the triplet solutions & \cite{Puntigam,HSocorro,electrovacMAG,M+S}&
 {\tt conom} $\sim$ {\tt
vecnom} $\sim$ {\tt trator}\\
\hline
\end{tabular}
\end{table}
\bigskip

\footnotesize
\noindent $^{*})$ Those pieces of the nonmetricity and the torsion vanish 
identically which are not mentioned in the description of a solution.
\normalsize

\clearpage

\noindent one is stuck, one changes the ansatz etc.

Beside the two dilation--shear solutions, Tresguerres
\cite{TresguerresShear1} and Tucker and Wang \cite{TuckerWang} found
Reissner--Nordstr\"om metrics together with a non--vanishing Weyl
covector, $^{(4)}Q^{\alpha\beta}\neq 0$, and a vector part of the
torsion, $^{(2)}T^\alpha\neq 0$, i.e., these solutions carry a {\em
  dilation} charge (in the words of Tucker and Wang, a Weyl charge)
and a {\em spin} charge, but are devoid of any other post--Riemannian
``excitations", in particular, they have no tracefree pieces
${\nearrow\!\!\!\!\!\!\!Q}_{\alpha\beta}$ of the nonmetricity.  As
shown by Tucker and Wang, the corresponding Lagrangian needs only a
Hilbert--Einstein piece $(a_0=1)$ and a segmental curvature squared
with $z_4 \neq 0$. The same has been proved for the Tresguerres
dilation solution, see footnote 4 of \cite{OVETH}.

Ho et al.\ \cite{Ho} found four spherically symmetric exact solutions
in a pure Weyl--Cartan spacetime which are similar to the dilation
type solutions. However, they include an additional axial part of the
torsion, $^{(3)}T^\alpha\neq 0$, see Table III.

\section{The triplet of post--Riemannian 1--forms and Obukhov's 
equivalence theorem}

The next step consisted in an attempt to understand the emergence of
the dilation--shear and the dilation--shear--torsion solutions of
Tresguerres.  However, as it so happened, it shifted the attention to
other types of solutions.  In both Tresguerres shear solutions, the
nonmetricity, besides the Weyl covector part {\tt conom}, was
represented by {\tt binom}, basically a 16 components' quantity.
However, {\tt conom} and {\tt trator} each have only 4 components, as
has {\tt vecnom}.  Accordingly, to create a simpler solution with
shear than the two Tresguerres dilation--shear solutions, it seemed
suggestive to require
\begin{equation}\label{triplet3} {\tt conom}\sim {\tt vecnom}\sim 
  {\tt trator}\,.
\end{equation}
This amounts to the presence of one 1--form $\phi$ which creates the
three post--Riemannian pieces (\ref{triplet3}). If $k_0$, $k_1$, $k_2$
are some constants (see below), then we have
\begin{equation}\label{triplet4} Q=k_0\phi\,,\quad \Lambda=k_1\phi\,,\quad T=
  k_2\phi\,,
\end{equation} with $\Lambda:=\vartheta^\alpha e^\beta \rfloor 
{\nearrow\!\!\!\!\!\!\!Q}_{\alpha\beta}$ and $T:=e_\alpha\rfloor
T^\alpha$.  This 1--form triplet was first proposed in
\cite{OVETH,VTOH} and also used in \cite {Dereli}.

Again, in the context of the triplet ansatz (\ref{triplet4}), a
Reissner--Nordstr\"om metric with a strong gravito--electric charge
could successfully be used \cite{OVETH} and a constraint on the
coupling constants had to be imposed.  Thus this ``triplet'' solution
is reminiscent of the Tresguerres dilation--shear solutions.  However,
its structure is simpler and, instead of {\tt binom}, it is {\tt
  vecnom} which enters the solution. Moreover, of the curvature square
pieces in the gauge Lagrangian $V_{\rm MAG}$ only the piece with
$z_4\neq 0$ is required. All others do not contribute.

Soon this result was generalized to an {\em axially symmetric}
solution \cite{VTOH} based on the {\em Kerr--Newman} metric and, a bit
later, to the whole {\em Pleba\'nski-Demia\'nski} class of metrics
\cite{PDMAG}. Already earlier, however, it became clear that the
triplet represents a general structure in the context of the
Einstein--Maxwellian ``seed'' metrics. In \cite{Dereli} it was pointed
out that for {\em each} Einstein--Maxwell solution (metric plus
electromagnetic potential 1--form), if the electric charge is replaced
by the strong gravito--electric charge and if a suitable constraint on
the coupling constants is postulated, an exact solution of MAG can be
created by means of the triplet (\ref{triplet4}). Even more so, if one
started from an {\em Einstein--Proca} solution instead, one could even
abandon the constraint on the coupling constants. This was first shown
for a certain 3-parameter Lagrangian by Dereli et al.\ \cite{Dereli}
and extended to a 6-parameter Lagrangian by Tucker \& Wang
\cite{TuckWar}.  The situation was eventually clarified for a fairly
general 11-parameter Lagrangian by the
\begin{itemize}
\item{} {\em Equivalence theorem of Obukhov \cite{Yuritheorem}:} Let
  be given the gauge Lagrangian $V_{\rm MAG}$ of (\ref{QMA}) with all
  $w_I=0$, $z_I=0$, except $z_4\neq 0$, i.e., the {\em segmental}
  curvature squared
\begin{equation}
  -\frac{z_4}{8\rho}\, R_\alpha{}^\alpha\wedge\,^\ast R_\beta{}^\beta=
  -\frac{z_4}{2\rho}\,dQ\wedge{}^\ast dQ\,\label{seg^2}
\end{equation} 
is the only surviving strong gravity piece in $V_{\rm MAG}$.  Solve
the Einstein--Proca equations\footnote{For the $\eta$--basis we have
  $\eta_{\alpha\beta\gamma}={}^\ast\left(\vartheta_\alpha\wedge
    \vartheta_\beta\wedge\vartheta_\gamma\right)$ and
  $\eta_\alpha={}^\ast\vartheta_\alpha$.}
\begin{eqnarray}\label{fieldob}
  \frac{a_0}{2}\,\eta_{\alpha\beta\gamma}\wedge\tilde{R}^{\beta\gamma}
  + \lambda \,\eta_\alpha &= \kappa\,\Sigma_\alpha^{(\phi)}\,,\\ 
  \left(\square+m^2\right)\phi &= 0\,,\\ \qquad d^\dagger\phi &= 0\,,
\end{eqnarray}
with respect to the metric $g$ and the Proca 1-form $\phi$.  Here the
tilde $\tilde{\null}$ denotes the Riemannian part of the curvature,
\begin{eqnarray}
  \Sigma_\alpha^{(\phi)}& := \frac{z_4k_0^2}{2\rho} \left\{ \left(
      e_\alpha\rfloor d\phi \right)\wedge{}^\ast d\phi- \left(
      e_\alpha\rfloor\,^\ast d\phi \right)\wedge\, d\phi \right.
  \nonumber \\ &\quad\left. +\;m^2\,\left[ (e_\alpha\rfloor
      \phi)\wedge{}^\ast \phi\;+\;
      (e_\alpha\rfloor\,^\ast\phi)\wedge{}\phi\right] \right\}\,
\label{ProcaEM}\end{eqnarray} 
is the energy--momentum current of the Proca field and $d^\dagger$ the
exterior {\em co}derivative. Then the {\em general vacuum solution} of
MAG with the stated parameter restrictions is represented by the
metric and the post--Riemannian triplet
\begin{equation}\left(g\,,\>\; Q=k_0\phi\,,\> \Lambda=k_1\phi\,,\> T=
k_2\phi\right)\,,                \label{MAGsolution}
\end{equation} where $k_0,k_1,k_2$ are elementary functions of the 
weak gravity coupling constants, $a_I,b_I,c_I$, and $m^2$ depends,
additionally, on $\kappa$ and the strong coupling constant $z_4/\rho$
(the details can be found in \cite{Yuritheorem}).
\end{itemize}
\noindent The results of \cite{Dereli,TuckWar} and of the Obukhov theorem 
lead to an understanding of the meaning of the constraint between the
different coupling constants: If we put $m^2=0$, then the
Einstein--Proca system becomes an Einstein--{\em Maxwell} system --
and such metrics, like the Kerr--Newman metric, e.g., are more readily
available for our purposes. In fact, we are not aware of any known
Einstein--Proca metrics which we could use for the construction of
exact MAG solutions. One should consult, however, the early work on
the Einstein-Proca system by Buchdahl \cite{Buch}, Ponomariov \&
Obukhov \cite{PonoObukhov}, and Gottlieb et al.\ \cite{Gott}. 

Also the reason for the more general character of the Tresguerres
shear solutions is apparent. He allowed gauge Lagrangians with
additional strong gravity pieces. In \cite{TresguerresShear1} he
added, to the segmental curvature piece, the strong gravity pieces
$w_3\times ({\tt pscalar})^2+w_5\times ({\tt ricanti})^2$. Here $(\;)^2$
is an abbreviation of $(\;)\wedge{}\! ^*(\;)$.  In this way he
circumvented the Obukhov theorem and found the spin 2 piece of the
nonmetricity, {\tt binom}, inter alia. On the other hand, the dilation
type solution in \cite{TresguerresShear1,TuckerWang} can be recovered
{}from the triplet solution \cite{OVETH} by means of a certain limiting
procedure, see \cite{OVETH}.

\section{Strong gravito--electric monopole, electrically charged 
versions of the triplet solutions}

In Table III, we gave an overview of the solutions of insular objects.
However, we didn't explain so far the first and the last entry of the
table.

The monopole type solution was found in \cite{Soliton2}, see also
\cite{Soliton1}, in terms of isotropic coordinates. In the Appendix we
translated the solution into Schwarzschild coordinates. Then, in these
coordinates, the orthonormal coframe, the metric, and the triplet
read, respectively,
\begin{equation}
  \vartheta ^{\hat{0}} =\,\left(1-\frac{q}{r}\right)\, d\,t \,,\quad
  \vartheta ^{\hat{1}} =\, \frac{d\, r}{1-\frac{q}{r}}\, , \quad
  \vartheta ^{\hat{2}} =\, r\, d\,\theta\,,\quad \vartheta ^{\hat{3}}
  =\, r\, \sin\theta \, d\,\varphi
\label{frame3}\,, 
\end{equation}  
\begin{eqnarray} g&=& \vartheta ^{\hat{0}}\otimes \vartheta ^{\hat{0}}-
\vartheta ^{\hat{1}}\otimes \vartheta ^{\hat{1}}-
\vartheta ^{\hat{2}}\otimes \vartheta ^{\hat{2}}-
\vartheta ^{\hat{3}}\otimes \vartheta ^{\hat{3}}\nonumber\\&=&
\left(1-\frac{q}{r}\right)^2dt^2-\frac{d\,r^2}
  {\left(1-\frac{q}{r}\right)^2}
  -r^2\left(d\,\theta^2+\sin^2\theta\,d\,\varphi^2 \right)\,,
\label{metric3}\end{eqnarray}
\begin{equation}\phi=\frac{Q}{k_0}=\,\frac{\Lambda}{k_1}=\,\frac{T}{k_2}=\,
  \frac{N_{\rm e}}{r\left(1-\frac{q}{r}\right)}\,\vartheta^{\hat{0}}=
  \frac{N_{\rm e}}{r}\,d\,t\,,
\label{monotrip1}
\end{equation}with $q=\sqrt{\frac{z_4\kappa}{2a_0\rho}}
\,k_0 N_{\rm e}$, i.e., it is again a triplet solution.

Note that the metric is {\em not} of the Schwarzschild form, the Weyl
covector, however, behaves as one expects for a strong gravito--{\em
  electric} charge. We recognize in this example in a particularly
transparent way that the strong gravito--electric charge $N_{\rm e}$
creates the post--Riemannian potentials {\tt conom}, {\tt vecnom},
{\tt trator} in (\ref{monotrip1}) in a quasi--Maxwellian fashion but
also emerges, in (\ref{metric3}), in the components of the metric.
However, in the metric, $N_{\rm e}$ behaves neither Schwarzschildian
(the metric is different) nor Reissner--Nordstr\"omian (the power of r
is reciprocal instead of $r^{-2}$).

We can construct this metric by a specific choice of the {\em mass} of
the Reissner-Nordstr\"om metric.\footnote{Private communications by
  D.\ Kramer (Jena) and M.\ Toussaint (Cologne).} In other words,
the metric of this solution represents a {\em subcase of the
  Reissner-Nordstr\"om metric}. Then it is immediately clear that this
solution is covered by the Obukhov theorem: One starts from an
Einstein--Maxwell solution, namely the Reissner--Nordstr\"om metric,
supplements the corresponding triplet, and chooses the mass such that
the Reissner-Nordstr\"om function $1-2m/r +q^2/r^2$ becomes a pure
square.

In the meantime, also a strong gravito-magnetic monopole has been
found \cite{M+S}. The mechanism is analogous to the gravito--electric
case and doesn't seem to bring new insight.

% 
% For the dilation--shear Lagrangian (\ref{dilsh}), we find 
% \begin{equation}
%   q^2=\kappa\frac{\alpha}{2}\,\left(k_0 N_{\rm e} 
% \right)^2\,,\quad
%   {\rm with}\quad k_0=-\frac{24}{\beta+6}\,,\>
%   k_1=-\frac{36\beta}{\beta+ 6}\,,\> k_2=6\,,
% \end{equation}
% \begin{equation} {\rm i.e.,}\qquad g= 
% \left(1-\frac{\sqrt{\frac
%         {\kappa\alpha}{2}}k_0 N_{\rm e} }{r}
% \right)^2dt^2+\cdots
% \end{equation}
% 
The last entry of Table III indicates that we are always able to find
electrically charged versions of a MAG solution as long as we confine
ourselves to the triplet type solutions. This is evident from
Obukhov's theorem: We take an electrically uncharged MAG solution with
the triplet $\sim\phi$. Then we choose the electromagnetic potential
$A$ proportional to the 1--form $\phi$.  Thus the structure of the
energy--momentum currents of the 1--form $\phi$ and the 1--form
$A$ is the same one. Both currents differ only by a constant.
Accordingly, they just add up, on the right hand side of the Einstein
equation, to a total energy--momentum current carrying a modified
constant in front of it.  Clearly, this structure breaks down as soon
as one turns to the full Einstein-Proca system, i.e., as soon as the
Proca mass becomes non--vanishing. Nevertheless, it is quite useful to
have found these electrically charged solutions explicitly. It helps
to illustrate the coupling of the electromagnetic field to the
post-Riemannian structures of a metric-affine spacetime, see
\cite{Puntigam}.

\section{Wave solutions}

{\em Plane--fronted} metric--Weyl covector--torsion {\em waves} have
been constructed by Tucker and Wang \cite{TuckerWang}. Their source is
a semi--classical Dirac spinor field $\psi(x)$. Let $\gamma^\alpha$ be
the Dirac matrices. Then the Dirac spin current
$\sim\overline{\psi}\gamma\gamma\gamma\psi$ generates the torsion
according to $\overline{\psi}\gamma\gamma\gamma\psi\sim$ {\tt tentor}
+ {\tt axitor}, whereas the Weyl covector and the torsion trace are
proportional to each other and are induced by the segmental curvature
square piece in the Lagrangian: {\tt conom} $\sim$ {\tt trator}. Thus
we have in this model an underlying Weyl--Cartan spacetime since the
tracefree part of the nonmetricity vanishes. In other words, the
solution is of the dilation type. Accordingly, the vacuum part of the
(weak and strong) gravitational field can be understood as a
degenerate triplet solution and again, as remarked in
\cite{TuckerWang}, it is straightforward to include a Maxwell field
with electric (and possibly magnetic) charge.

In view of \cite{TuckWar} and the Obukhov theorem, it is clear that
one may start with any solution of the Einstein--Maxwell equations.
Then one replaces, after imposing a suitable constraint on the
coupling constants, the electric charge by the strong gravity charge
thereby arriving at the post--Riemannian triplet which was mainly
discussed in Sec.V. The procedure is fairly straightforward.
Nevertheless, it is useful to have a couple of worked--out examples at
one's disposal. Explicit solutions may convey a better understanding
of the structures involved.

Garcia et al.\ \cite{collwavesMAG} studied {\em colliding waves} with
the corresponding metric and an excited post--Riemannian triplet in
the framework of a Lagrangian of the Obukhov theorem.  Usually, in
general relativity, the colliding waves are generated by {\em
  quadratic polynomials} in the appropriate coordinates. And these
polynomials were also used in the paper referred to. Recently,
however, Bret\'on et al.\ \cite{collwavesGR} were able, within general
relativity, to extend this procedure by using also {\em quartic}
polynomials. Again, this procedure can be mimicked in metric--affine
spacetime and Garcia et al.\ \cite{shock5,Oscht} constructed
corresponding colliding gravity waves with triplet excitation. For the
quadratic as well as for the quartic case it is also possible to
generalize to the {\em electrovac} case, as has been shown in
\cite{electrovacMAG}.

\section{Cosmological solutions}

As we argued above, we expect more noticeable deviations from
metric--compatibility the further we go back in time. Therefore it is
natural to investigate cosmological models in the framework of
metric--affine gravity. And the standard {\em Friedmann} model is a
good starting point. Tresguerres \cite{Tres3} proposed such a model
with torsion and a Weyl covector, i.e., spacetime is described therein
by means of a Weyl--Cartan geometry. The matter he used to support the
model is a fluid carrying an energy--momentum and a dilation current.
The field equations of the model stayed within a manageable size since
the Lagrangian, by assumption, carries only the segmental curvature
square piece of the {\em symmetric} part of the curvature 2--form.
However, the square of all 6 irreducible pieces of the {\em
  anti}symmetric part of the curvature are allowed in the
gravitational Lagrangian even if only the tracefree symmetric Ricci
turns out to be relevant in the end. A somewhat similar model has been
investigated by Minkevich and Nemenmann \cite{Minkevich}.

Using the much more refined model of a hyperfluid \cite{Ob2}, Obukhov
et al.\ \cite{Yuritheorem} derived, within the framework of the the
equivalence theorem, but with some additional simplifying assumptions,
a Friedmann cosmos with a time varying Weyl covector. This is
analogous as in the Tresguerres model.

Similar structures have been suggested by Tucker and Wang
\cite{TuckerWang2}. They proposed a metric--affine geometry of
spacetime for the purpose of taking care of the supposedly unseen dark
matter which, as they suggest, interacts with the strong gravity
potential of the Proca type as described by means of a gravitational
Lagrangian carrying a segmental curvature square. Thus the Obukhov
theorem applies to their scenario, and a Friedmann solution with a
post--Riemannian triplet is expected to emerge. And this is exactly
what happens. Ordinary matter and dark matter both supply their own
material energy--momentum current to the right hand side of the
Einstein equation and, additionally, a Proca energy-momentum comes up,
see (\ref{fieldob},\ref{ProcaEM}). The material current that couples
to the Proca field can be identified with the trace of the material
hypermomentum current, the material dilation current, see the trace of
the right hand side of (\ref{second}). The model is worked out in
considerable detail, galactic dynamics and the cosmological evolution
are studied inter alia and numerical results presented.

\section{The minimal dilation--shear Lagrangian, ansatz with a Proca 
`mass'}

Taking the triplet (\ref{triplet4}) as a guide, it is certainly helpful
for model building not to take the whole weak part of (\ref{QMA}) but
only some sort of essential nucleus of it. Putting (\ref{4}) and
(\ref{5}) together, one gets certainly a propagating Weyl covector.
{}From the Obukhov theorem we know that we only need a further weak
gravity piece in order to allow for shear. In view of the triplet, the
addition of a {\tt trator} square piece is suggested.  In this way we
recover the minimal dilation--shear Lagrangian \cite{OVETH,Dereli}
\begin{equation}
  V_{\rm dil-sh} = \frac{1}{2\kappa} \left( - R^{\alpha\beta} \wedge
    \eta_{\alpha\beta} + \beta\, Q \wedge {}^*Q+ \gamma\, T \wedge
    {}^*T\right)- \frac{\alpha}{8}\,R_\beta{}^\beta \wedge
  {}^*R_\gamma{}^\gamma
\label{dilsh}\, .
\end{equation}
And indeed, our Reissner--Nordstr\"om, Kerr--Newman, and
Pleba\'nski--Demia\'nski metrics, together with the post--Riemannian
triplet (\ref{triplet4}), with the constants
\begin{equation}k_0=-\frac{3}{2}\,\gamma-4\,,\qquad 
                k_1=\frac{27}{2}\,\gamma\,,\qquad k_2=6\,,
\end{equation}and with the 1--form ($N$ is an integration constant)
\begin{equation}\phi=\frac{N}{r}\,d\,t\,,\end{equation}
are solutions of the field equations belonging to the $V_{\rm
  dil-sh}$ Lagrangian. However, a constraint on the weak coupling
constants has to be imposed:
\begin{equation}\label{conx1}\gamma=-\frac{8}{3}\, \frac{\beta}{\beta+6}\,.
\end{equation}
Accordingly, the Lagrangian (\ref{dilsh}) may be considered as the
generic Lagrangian of the Obukhov theorem.

Let us now try to get rid of the constraint (\ref{conx1}). The
corresponding procedure runs as follows: According to
\cite{Yuritheorem} Eq.\ (6.8), we can define the Proca mass
\begin{equation}m_{\rm Proca}^2=\frac{1}{2\kappa\alpha}\,\left(2\beta+
\frac{36\gamma}{3\gamma+8}\right)\,.
\end{equation}
If we put it to zero, we recover the constraint (\ref{conx1}):
\begin{equation}m_{\rm Proca}^2=0\quad\longrightarrow\quad
\gamma=-\frac{8}{3}\, \frac{\beta}{\beta+6}\,.
\end{equation} Thus the dropping 
of the constraint (\ref{conx1}) is equivalent to the emergence of a
Proca mass, i.e., we now have to turn to the Einstein--Proca system
instead of to the Einstein--Maxwell system.

Then, in {\em flat} spacetime, after dropping the constraint
(\ref{conx1}), instead of a Coulomb potential, we expect a Yukawa
potential to arise as a solution of the Proca equation:
\begin{equation}\phi\sim N\,\frac{e^{-m_{\rm Proca}r}}{r}\,d\,t\,.
\end{equation}In the corresponding metric--affine spacetime, the 
Reissner-Nordstr\"om metric has also to be modified. If done in a
suitable way, this should lead to an exact solution of the {\em
  unconstrained} dilation--shear Lagrangian (\ref{dilsh}).

%*****************************************************
\section{Discussion}

In the last section we have already seen, how we can hope to extend
our work. But also a generalization in another direction is
desirable. If we want to include the shear solutions of Tresguerres,
then the dilation--shear Lagrangian is too narrow. To go beyond the
triplet solution requires a generalization of (\ref{dilsh}). A `soft'
change, by switching on only the post-Riemannian pieces of the {\em
  antisymmetric} piece of the curvature 2--form, seems worth a try:
\begin{equation}V_{\rm dil-sh-tor}\sim V_{\rm dil-sh} -\frac{1}{2\rho}\left[
    w_2\times ({\tt paircom})^2+w_3\times({\tt
      pscalar})^2+w_5\times({\tt ricanti})^2 \right]\,.\end{equation}
A related model was discussed in \cite{Bianchi} Sec.5.3. In this way
we can hope to `excite', besides {\tt conom} and {\tt vecnom}, also
{\tt binom}, e.g. Of course, also in this case one should try to
remove the constraint. However, it will not be sufficient in this
case, as is clear from \cite{Yuritheorem} and the Obukhov equivalence
theorem, to turn only to the Einstein--Proca system --- rather a more
general procedure will be necessary.

%******************************************************
\section{Appendix: Strong gravito-electric monopole in 
  Schwarzschild coordinates} 

In \cite{Soliton2}, the MAG solution of the soliton type was given in
terms of isotropic coordinates. This makes it more difficult to
compare it with the Reissner--Nordstr\"om type solution. Therefore we
will perform a coordinate transformation. We will denote the isotropic
polar coordinates by $(t,\rho,\theta,\varphi)$ and the Schwarzschild
coordinates by $(t,r,\theta,\varphi)$.  In \cite{Soliton2}, the
following monopole solution has been found: The orthonormal coframe
reads
\begin{equation}
  \vartheta ^{\hat{0}} =\,{1\over f}\, d\,t \,,\quad
\vartheta ^{\hat{1}} =\, f\, d\, \rho\, , \quad
\vartheta ^{\hat{2}} =\, f\, \rho\, d\,\theta\,,\quad
\vartheta ^{\hat{3}} =\, f\, \rho\, \sin\theta \, d\,\varphi
\label{frame2}\, ,
\end{equation}
with the function 
\begin{equation} f(\rho)= 1+\frac{q}{\rho}\,,
\label{monopole}\end{equation}
and the one--form triplet is specified by (in this Appendix, $\rho_{\rm
  c}$ denotes the strong gravity coupling constant)
\begin{equation}\phi=
  \frac{Q}{k_0}=\,\frac{\Lambda}{k_1}=\,\frac{T}{k_2}=\, \frac{N_{\rm
      e}}{\rho}\,\vartheta^{\hat{0}}\,,\qquad{\rm with}\qquad
  {q}^2=\frac{z_4\kappa}{2a_0\rho_{\rm c}}\,\left(k_0N_{\rm
      e}\right)^2
\label{monotrip}\, .
\end{equation}

For the transition to Schwarzschild coordinates, the
$\theta$--component of the coframe has to obey
\begin{equation} \vartheta^{\hat{2}}=\left(1+\frac{q}{\rho}\right)\rho
  \,d\theta=r\,d\theta\,.\end{equation} Thus
\begin{equation} r=\left(1+\frac{q}{\rho}\right)\rho=\rho+q\,,\qquad
dr=d\rho\,.\end{equation}
Substitution into (\ref{monopole}) yields
\begin{equation}f=1+\frac{q}{\rho}=\frac{r}{\rho}=\frac{r}{r-q}=
  \frac{1}{1-\frac{q}{r}}\,.\end{equation} Accordingly, the monopole
solution can be rewritten in the form as displayed in
(\ref{frame3},\ref{metric3},\ref{monotrip1}).

\acknowledgments

This research was supported by the joint German--Mexican project
DLR--Conacyt E130--2924 and MXI 009/98 INF and by the Conacyt grant
No.\ 28339E. FWH would like to thank Alfredo Mac\'{\i}as and the
Physics Department of the UAM--I for hospitality.  Furthermore, we
appreciate helpful remarks by Tekin Dereli (Ankara), Dietrich Kramer
(Jena), Jim Nester (Chung-li), Yuri Obukhov (Moscow), Jos\'e Socorro
(Le\'on), Marc Toussaint (Cologne), Robin Tucker (Lancaster), and
Charles Wang (Lancaster).
%********************************************************

%\end{thebibliography}

%%-->
%\end{multicols}
%%--
\end{document}